\documentclass[10pt,twocolumn,letterpaper]{article}

\usepackage[pagenumbers]{cvpr} % To force page numbers, e.g. for an arXiv version

\definecolor{cvprblue}{rgb}{0.21,0.49,0.74}
\usepackage[pagebackref,breaklinks,colorlinks,allcolors=cvprblue]{hyperref}
\usepackage{bm}
\usepackage{colortbl} % For cell coloring
\usepackage{array}    % For custom column width
\usepackage{graphicx} % For adjusting table spacing
\usepackage{mathabx}
\usepackage{wrapfig}
\newcommand\blfootnote[1]{%
\begingroup
\renewcommand\thefootnote{}\footnote{#1}%
\addtocounter{footnote}{-1}%
\endgroup
}

\newcommand{\METHOD}{PhysAnimator}

\def\paperID{2235} % *** Enter the Paper ID here
\def\confName{CVPR}
\def\confYear{2025}

\title{\METHOD: Physics-Guided Generative Cartoon Animation}

\author{Tianyi Xie$^{1,2*}$  \quad Yiwei Zhao$^{1\text{\textdagger}}$\quad Ying Jiang$^{2}$ \quad Chenfanfu Jiang$^{2}$ \\
$^{1}$ Netflix, $^{2}$ UCLA \\
{\tt\small tianyixie77@ucla.edu} \quad {\tt\small yiweiz@netflix.com} \quad
{\tt\small \{yingjiang, cffjiang\}@ucla.edu}
}

\begin{document}
% \begin{center}
%     \centering
%     \captionsetup{type=figure}
%     \includegraphics[width=\textwidth, trim=0 0 0 0, clip]{images/physanimeteaser.pdf}
%     \captionof{figure}{GarmentDreamer is a garment synthesis framework for customizing simulation-ready high-quality textured garment meshes from text prompts.}
% \label{fig:teaser}
% \end{center}
\twocolumn[{%
\renewcommand\twocolumn[1][]{#1}%
\maketitle
\begin{center}
    \centering
    \captionsetup{type=figure}
    \includegraphics[width=\textwidth, trim=0 0 0 0, clip]{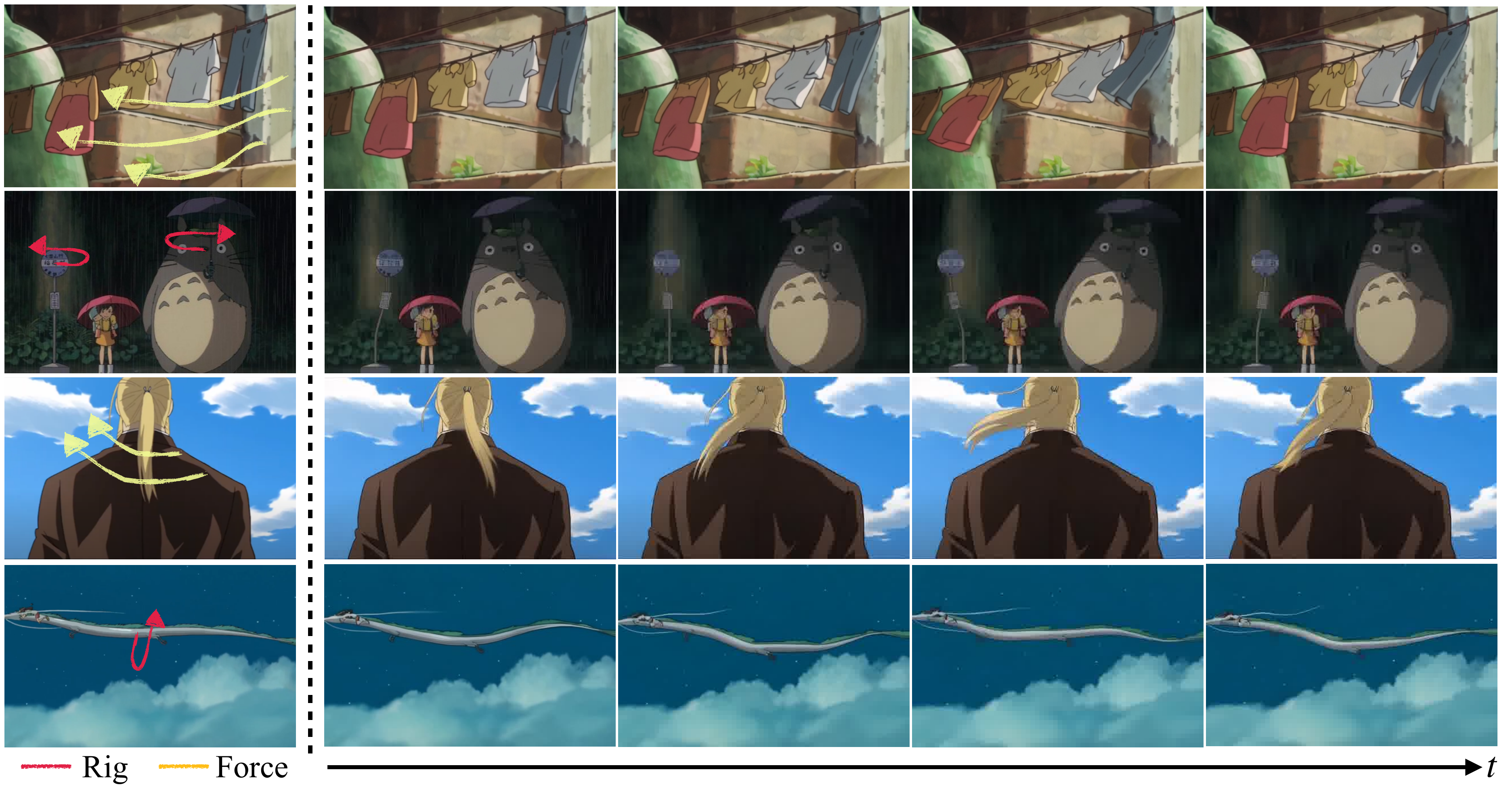}
    \vspace{-15px}
    \captionof{figure}{\textbf{\METHOD} is a novel framework that combines physics principles with video diffusion models to generate high-quality animations from static anime illustrations, allowing users to specify external forces or rigging points for custom effects.}
\label{fig:teaser}
\end{center}
}]
% \maketitle
\begin{abstract}
\blfootnote{* Work done during the internship at Netflix.} 
\blfootnote{\textdagger\ Corresponding author.}
Creating hand-drawn animation sequences is labor-intensive and demands professional expertise. We introduce \METHOD, a novel approach for generating physically plausible meanwhile anime-stylized animation from static anime illustrations. Our method seamlessly integrates physics-based simulations with data-driven generative models to produce dynamic and visually compelling animations. To capture the fluidity and exaggeration characteristic of anime, we perform image-space deformable body simulations on extracted mesh geometries. We enhance artistic control by introducing customizable energy strokes and incorporating rigging point support, enabling the creation of tailored animation effects such as wind interactions. Finally, we extract and warp sketches from the simulation sequence, generating a texture-agnostic representation, and employ a sketch-guided video diffusion model to synthesize high-quality animation frames. The resulting animations exhibit temporal consistency and visual plausibility, demonstrating the effectiveness of our method in creating dynamic anime-style animations. See our project page for more demos: \url{https://xpandora.github.io/PhysAnimator/}.
\end{abstract} 
\section{Introduction}
\label{sec:intro}

Dynamic visual effects are essential to the immersive quality of 2D animation. From the subtle sway of a character’s hair to the fluid motion of garments in response to wind, realistic and pleasing dynamics create a captivating visual experience. Traditionally, these effects are achieved through meticulous, hand-drawn techniques, where animators painstakingly draw each frame to bring these dynamic elements to life. This process is labor-intensive and requires not only artistic skill but also a deep understanding of natural forces and environmental interactions.

To alleviate the challenges of manual animation, researchers have explored both traditional and data-driven approaches. Traditional animation tools \cite{kazi2016motion, su2018live} provide interactive systems that assist artists in creating animated illustrations based on established principles of 2D animation. These methods often rely on user-provided stroke inputs to specify motion trajectories and utilize geometry constraints to produce deformation-based animation. However, such approaches typically assume simple inputs like lineart or drawings with separated layers, limiting their applicability to more complex, in-the-wild anime illustrations that feature intricate textures and details. In contrast, data-driven video generative models \cite{xing2025dynamicrafter, blattmann2023stable, ma2024cinemo} offer a promising alternative by leveraging neural synthesis to generate dynamic effects directly from images, bypassing the need for manual sketching or layered inputs. Recent methods even enable interactive motion design, allowing users to specify object trajectories via drag-based inputs \cite{wu2025draganything, niu2024mofa, shi2024motion}. These approaches typically rely on predicting a sequence of optical flow fields to drive the desired motion and warp the frames accordingly. However, the quality of the generated results is often limited by inaccuracies in the predicted optical flow, which frequently exhibits artifacts due to a lack of geometric understanding and physical constraints. As a result, imprecise motion estimation can lead to noticeable distortions and unsatisfactory visual quality.

Recognizing the limitations of both traditional and purely data-driven methods, we introduce \METHOD, a novel framework that integrates physics-based animation with data-driven generative models to synthesize visually compelling dynamic animations driven by environmental forces such as wind or rigging from static anime illustrations. Our approach combines the physical consistency of simulation-based methods with the flexibility and expressiveness of generative models, overcoming the drawbacks of previous approaches. To achieve this, we model objects of interest in the anime illustration as deformable bodies, capturing the fluidity and exaggerated motion characteristic of anime. We solve the motion equations to compute a sequence of optical flow fields that represent consistent dynamics. Users can enhance these animations with customized energy strokes, defining the effects of external forces such as wind. To render the motion dynamics as high-quality frames, we first extract and warp the sketch using the optical flow fields to generate a texture-agnostic video sequence. We then apply a sketch-guided video diffusion model to colorize the sketch sequence based on the reference illustration, ensuring stylistic coherence. Finally, to incorporate dynamic effects that cannot be easily captured by physical simulation, we employ a data-driven cartoon interpolation model, enriching the results with complementary, expressive dynamics. In summary, our contributions are as follows:
\begin{itemize}
    \item We introduce a novel framework that combines physics-based simulations with data-driven generative models, specifically targeting animations driven by environmental forces and rigging controls, achieving both physical consistency and stylistic expressiveness.
    \item We develop an image-space deformable body simulation technique that models anime objects as deformable meshes, capturing fluid and exaggerated dynamics.
    \item  We leverage a sketch-guided video diffusion model to render simulation dynamics as high-quality frames, combined with a cartoon interpolation model to introduce additional dynamic effects beyond physical laws.
\end{itemize}

% In this work, we propose a novel approach that combines the strengths of physics-based simulations with data-driven generative models to synthesize visually compelling, physics-consistent dynamics from an anime image.
\section{Related Work}
\label{sec:related}

\subsection{Video Diffusion Models}
Building on the success of image diffusion models \cite{dhariwal2021diffusion, rombach2022high}, recent work has introduced video diffusion models to streamline video synthesis from text prompts or images, significantly reducing labor and time costs compared to traditional commercial video editing and creation tools \cite{ho2022video, xing2023survey}. These approaches either extend pre-trained image diffusion models by incorporating temporal mixing layers in various forms \cite{singer2022make, blattmann2023align, ge2023preserve, esser2023structure} or train video diffusion models with temporal layers from scratch on large-scale text-video paired datasets \cite{he2022latent, zhang2024trip}. However, using text prompts or a single image as input provides limited control over fine-grained aspects of video generation, making it challenging to define complex structural attributes such as spatial layouts, poses, and shapes \cite{guo2025sparsectrl, yin2023dragnuwa}. To allow for more detailed control over object motion and camera movements, additional guidance in the form of motion trajectories \cite{jeong2024vmc, shi2024motion, yin2023dragnuwa, chen2023control, tan2024physmotion}, pose \cite{zhang2024mimicmotion}, depth \cite{he2023animate}, and optical flow \cite{guo2025sparsectrl} has been integrated into video diffusion models to produce more controllable videos. These powerful video diffusion models have also been applied to various downstream tasks, such as video editing \cite{ku2024anyv2v, gu2024videoswap}, image animation \cite{gal2024breathing, xing2025dynamicrafter, wu2024aniclipart}, video understanding \cite{wang2024zero, lee2024video, shao2024learning}, video interpolation \cite{xing2024tooncrafter, jain2024video} and 3D reconstruction and generation \cite{voleti2025sv3d, han2025vfusion3d, kwak2024vivid, chen2024v3d, liu2024reconx}. Nevertheless, these data-driven approaches usually produce artifacts due to a lack of geometric understanding and physical constraints \cite{bansal2024videophy, he2024videoscore}.

\subsection{Physics-based Animation}
The animation authoring pipeline includes stages such as sketching, modeling, rigging, and animation \cite{jiang2021handpainter, jiang2024region, liu2025riganything, maestri2006digital}. Among these, the creation of dynamic visual effects plays a crucial role in enhancing the immersive quality of 2D animation. 
%Researchers have explored various methods for animation authoring. 
\citet{xing2015autocomplete} extend local similarity techniques to global similarity, enabling automatic sketch completion based on previous frames. Building on this, \citet{peng2020autocomplete} propose a keyframe-based sculpting system that autocompletes user edits using an intuitive brushing interface. \citet{willett2018mixed} tackle the challenge of textured anime images by segmenting the illustration into layers for different objects and allowing users to provide scribbles as trajectory guides. Leveraging animation principles, some works \cite{ma2022layered, kazi2016motion} use geometry-based deformations to create stylized animation effects such as ``squash" and ``stretch". Similarly, \citet{xing2016energy} employ physics-based simulations and introduce energy brushes to generate elemental dynamics \cite{gilland2012elemental}, such as smoke and fire. Other approaches \cite{coros2012deformable, jones2015dynamic, barbivc2009deformable} incorporate physical laws like gravity, collision, and elasticity to animate deformable characters, enabling the efficient creation of complex scenes.  For instance, \citet{jones2016example} introduces an example-based plasticity model based on linear blend skinning for animating the failure of near-rigid, man-made materials. \citet{coros2012deformable} exploit two rest-pose adaptation strategies using only internal energy to animate curve, shell, and solid-based characters. Physics-based modeling has also proven effective in adding secondary dynamics \cite{willett2017secondary, zhang2020complementary} to enhance details in rigged animations. Despite their effectiveness, these 2D animation methods typically assume simple inputs, such as sketches, or require layered separation, which limits their applicability for animating complex, in-the-wild anime illustrations.

\subsection{Generative Image Dynamics}  
%Image Dynamics has two diagrams: 1. data-driven: rely on generative models, lacking understanding physics n 2. simulators: 2d/3d image dynamics; we focus on 2d dynamics without generate 3d contents; time consuming; expensive; for waving effects
With the advancement of generative models, there is growing interest in leveraging these powerful techniques to synthesize dynamic animations from static images, guided by motion features extracted from various user inputs. These inputs can be sparse, such as text prompts \cite{xing2025dynamicrafter, chen2023seine, li2024animate}, trajectories \cite{wang2024motionctrl, wu2025draganything, shi2024motion, zhang2024tora, li2024generative}, or camera movements \cite{wang2024motionctrl, yang2024direct}, or dense, like reference videos \cite{wei2024dreamvideo, zhao2023motiondirector, jeong2024vmc}. To achieve controllable dynamics, prior works often incorporate ControlNet \cite{zhang2023adding} into image or video generative models during the decoding stage, utilizing motion features such as Canny edges, depth maps \cite{guo2025sparsectrl}, 2D Gaussian maps \cite{wu2025draganything}, and optical flow maps \cite{shi2024motion}. In addition, several methods employ specialized motion fusion modules \cite{zhang2024tora, wang2024motionctrl, li2024animate}, spatial or temporal attention mechanisms \cite{yang2024direct, jeong2024vmc, wei2024dreamvideo}, and feature injection strategies \cite{li2024generative} to guide the generation of controllable dynamics. However, these approaches typically lack physics-based supervision, which can result in animations that violate physical laws or fail to align with user intentions \cite{meng2024towards}. Recently, PhysGen \cite{liu2025physgen} integrated a rigid-body physics simulator to generate physics-consistent dynamics for foreground objects in a given image. While effective, PhysGen is limited to 2D rigid-body motions, making it unsuitable for the fluid, elastic effects commonly seen in anime, such as the waving of cloth or hair, which do not conform to rigid-body dynamics. Recent works like VR-GS \cite{jiang2024vr}, PhysGaussian \cite{xie2024physgaussian}, and PhysDreamer \cite{zhang2025physdreamer} employ physics-based simulators and differentiable renderers to synthesize high-quality videos, but require multi-view images as inputs. In this work, we address these limitations by leveraging deformable-body dynamics and a sketch-guided video diffusion model to create high-quality, physically consistent animation sequences from a single image, effectively capturing the fluid and elastic motion typical of anime.
% To create physics-grounded image dynamics, we utilize XPBD \cite{macklin2016xpbd} simulators to generate intermediate control signals as input to video diffusion models. 
% Recent work, PhysGen \cite{liu2025physgen}, also utilizes rigid-body physics simulator to create physics-based dynamics but it is limited to 2D rigid-body motions.  While dynamics in anime encompass not only rigid-body motions but also involve elasticity, fluidity, waving effects of cloth and hair, etc. \cite{xing2016energy, kazi2016motion}. \TODO{to be revised}
% In summary, our work focuses on generating visually compelling, anime-style, physics-grounded dynamics from a single image.
\section{Method}
\begin{figure*} [t]
 \includegraphics[width=1.0\textwidth]{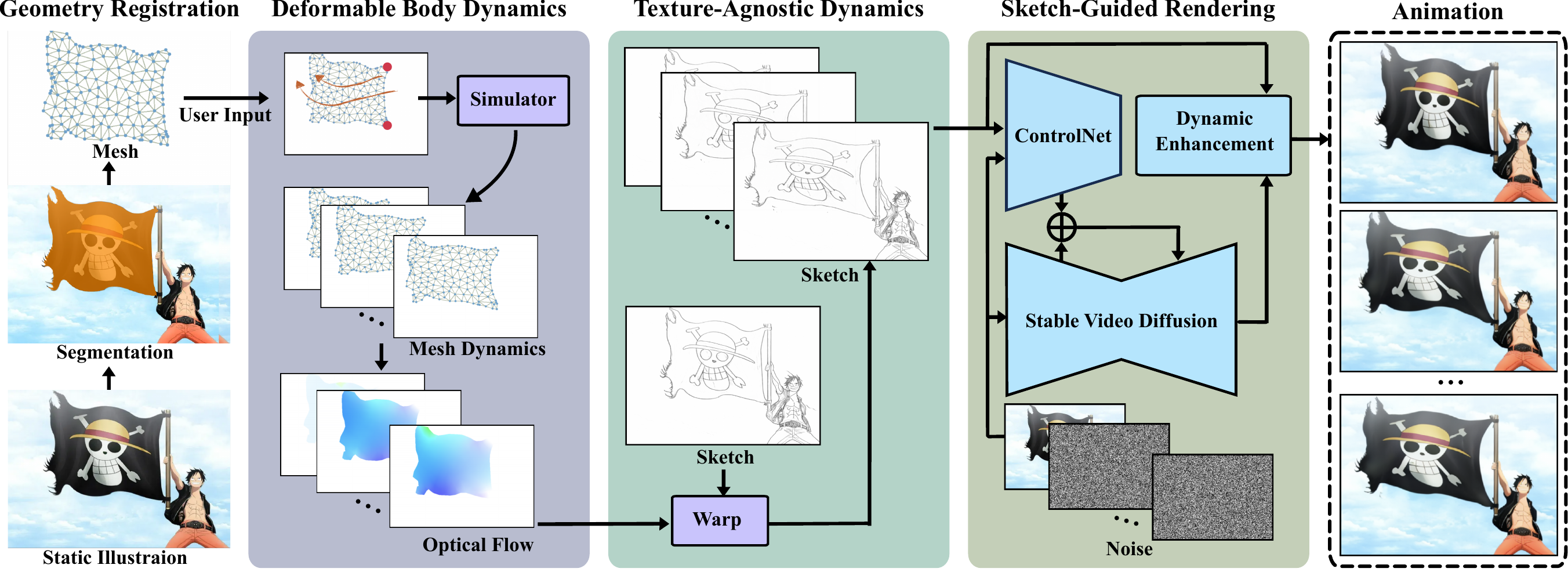}
  \centering
    \caption{\textbf{Method Overview.} We begin by segmenting the object and creating a triangulated deformable mesh. Physics-based simulations are then used to generate dynamic optical flow fields, with users given the option to guide the motion through customizable energy strokes(shown as orange arrows) and rigging points(shown as red dots). The extracted sketch is warped using the computed optical flow and refined with a sketch-guided video diffusion model, producing a smooth, stylized animation sequence. Optionally, a cartoon interpolation model can further be applied to enhance the animation with expressive dynamics.}
  \label{fig:pipeline}
  \vspace{-4mm}
\end{figure*}

% Given a reference anime illustration $I_0$, our goal is to generate a sequence of stylized video frames $\{\hat{I}_1, \hat{I}_2, ..., \hat{I}_T\}$ with natural, user-guided motions. To achieve this, we propose \METHOD, a novel framework that combines video diffusion with physics-based animation. We begin by segmenting and triangulating the objects of interest to create a mesh of the illustration. With the obtained geometry, we generate smooth dynamics with physics-based simulation, producing a sequence of optical flow fields $\{\mathcal{F}_{0 \rightarrow 1}, \mathcal{F}_{0 \rightarrow 2}, ..., \mathcal{F}_{0 \rightarrow T}\}$. Users can interactively apply energy strokes to define external forces or specify rigging points for desired animation effects. In the final generative rendering stage, we extract the sketch from the input illustration and warp it with the optical flow fields, creating a sequence of dynamic sketch frames $\{S_1, S_2, ..., S_T\}$. These sketches and the input illustration are then fed into the stable video diffusion model with a sketch-guided ControlNet, rendering a vivid animated video sequence. We further utilize a data-driven cartoon interpolation model to enhance the results with complementary dynamics that go beyond physics-laws. Fig.~\ref{fig:pipeline} provides an overview of our proposed framework.
Given a reference anime illustration $I_0$, our goal is to generate a sequence of stylized video frames $\{\hat{I}_1, \hat{I}_2, ..., \hat{I}_T\}$ with natural, user-guided motions. To achieve this, we introduce \METHOD, a novel framework that combines video diffusion models with physics-based animation. We start by segmenting the objects of interest in a given anime illustration and creating a 2D triangulated mesh representation. Using this geometry, we obtain dynamics through image-space deformable body simulation, generating a sequence of optical flow fields $\{\mathcal{F}_{0 \rightarrow 1}, \mathcal{F}_{0 \rightarrow 2}, ..., \mathcal{F}_{0 \rightarrow T}\}$. To offer user control, we enable interactive inputs through customizable energy strokes for defining external forces and rigging points for specifying desired trajectories. To render dynamics as high-quality frames, we extract the sketch from the input illustration and warp it using the computed optical flow fields, yielding a sequence of dynamic sketch frames $\{S_{1}, S_{2}, ..., S_{T}\}$. These sketches, together with the input illustration, are fed into a video diffusion model with a sketch-guided ControlNet \cite{zhang2023adding}, synthesizing a vivid animated sequence. Finally, an optional data-driven cartoon interpolation model \cite{xing2024tooncrafter} can be applied to enhance the anime style dynamics of the results by selecting keyframes from the animated sequence as input, yielding the final output frames $\{\hat{I}_1, \hat{I}_2, ..., \hat{I}_T\}$. An overview of our proposed framework is shown in Fig.~\ref{fig:pipeline}.
% To further enhance the \yiwei{smoothness and anime the style of the} results, we apply a data-driven cartoon interpolation model \yiwei{between the animated key frames}. This introduces complementary dynamics that extend beyond physical simulations, thus generating the final out frames $\{\hat{I}_1, \hat{I}_2, ..., \hat{I}_T\}$. An overview of our proposed framework is shown in Fig.~\ref{fig:pipeline}.

\subsection{Preliminaries}
\vspace{-0mm}
\paragraph{Latent Video Diffusion Model}
Most video diffusion models build on the latent diffusion model (LDM) framework \cite{rombach2022high}, which uses Variational Autoencoder \cite{kingma2013auto} (VAE) to map input images into a latent space. In this space, data is transformed into Gaussian noise via a forward diffusion process, which the model learns to reverse through denoising. In the forward diffusion process, the latent code $z_0$ is perturbed as:
\begin{equation}
    z_t = \sqrt{\Bar{\alpha}_t} z_0 + \sqrt{1 - \Bar{\alpha}_t} \epsilon, \epsilon \sim \mathcal{N}(0, I),
\end{equation}
where $\Bar{\alpha}_t = \prod \limits_{i=1}^t(1 - \beta_t)$ with $\beta_t$ controlling the noise strength coefficient at step $t$. The denoising model $\epsilon_\theta$ is trained to recover $z_t$ by minimizing the objective function:
\begin{equation}
    L_\epsilon = \| \epsilon - \epsilon_\theta(z_t; c, t)\|^2_2,
\end{equation}
where $\theta$ denotes the learnable network parameters and $c$ represents conditioning input (e.g. text prompts or images). After denoising, the VAE decoder reconstructs the latent code back into the image space. The latent video diffusion model (LVDM) extends the image LDM to videos by incorporating temporal modules to maintain temporal consistency.
\vspace{-4mm}
\paragraph{Deformable Body Dynamics}
Mathematically, the dynamics of a continuum deformable body is described by a time-dependent continuous deformation map $\bm{x} = \phi(\bm{X}, t)$, which maps the undeformed material space $\Omega^0$ to the deformed world space $\Omega^t$ at time $t$. The deformation gradient $\bm{F} = \frac{\partial \phi}{\partial \bm X}$ encodes the local deformation such as scaling, rotation, and shearing. In the context of a discretized setting, e.g. 2D meshes, the deformation map for each triangle can be expressed as
\begin{equation}
    \phi_i(\bm{X}) = \bm{F}_i \bm{X} + \bm{b}_i,
\end{equation}
where $\bm{b}_i \in \mathbb{R}^2$ accounts for translation of $i$-th triangle, and the deformation gradient $\bm{F}_i \in \mathbb{R}^{2\times 2}$ is assumed to be constant \cite{li2024physics}. Undergoing deformation, the deformed body aims to recover its rest shape via resisting forces. Continuum mechanics model this behavior by first defining an energy density function $\Psi(\bm{F}(\bm{x}))$, which measures the strain energy per unit undeformed volume. The total potential energy for a deformable body is then obtained as
\begin{align}
    E(\bm{x}) = \sum \limits_{i=1}^N \Psi(\bm{F}_i) V_i,
\end{align}
where $V_i$ denotes volume of $i$-th triangle. The internal resisting force is then defined as the negative gradient of the potential energy with respect to the vertex position
\begin{equation}\label{eq: internal_forces}
    \bm{f}_{\text{int}}(\bm{x}) = -\frac{\partial E(\bm{x})}{\partial \bm{x}}.
\end{equation}
The deformable body dynamics is governed by Newton's Second Law as 
\begin{equation} \label{eq: time_integration}
    \frac{d^2 \bm{x}}{dt^2} = \bm{M}^{-1}(\bm{f}_{\text{int}}(\bm{x}) + \bm{f}_{\text{ext}}(\bm{x})),
\end{equation}
which can be solved numerically. Here $M$ is the mass matrix that represents the masses of all vertices and $\bm{f}_{\text{ext}}$ refers to external forces, such as gravity.
% one can solve the deformable body dynamics with a numerical time-stepping method, e.g. semi-implicit Euler:
% \begin{align}
%     \bm{x^{n+1}} &= \bm{x^n} + \Delta t \bm{v}^{n+1}, \\
%     \bm{v^{n+1}} &= \bm{v^n} + \Delta t \bm{M}^{-1} (\bm{f}_{\text{int}} + \bm{f}_{\text{ext}}),
% \end{align}
% where $\Delta t$ denotes the timestep size between $t^{n+1}$ and $t^n$ and $M$ represents the mass of all vertices. $\bm{f}_{\text{int}}$ and $\bm{f}_{\text{ext}}$ refer to internal and external forces respectively.

\subsection{Physics-based Animation}
To generate dynamic animations from a single illustration, a straightforward approach is to use existing controllable video diffusion models \cite{wu2025draganything, shi2024motion}. However, these purely data-driven methods often struggle due to a lack of physical understanding, leading to unrealistic results. To overcome this limitation, we incorporate physics-based animation to generate physically consistent and plausible motions.

% To create dynamics for a single illustration image, one straightforward way is to leverage controllable video diffusion models. To enable \yiwei{positional} motion control \yiwei{other than text} in video diffusion models, current approaches \cite{niu2024mofa, shi2024motion} typically rely on a user-defined trajectory to guide optical flow predictions through a dedicated network \cite{zhan2019self}. These predicted optical flow fields serve as additional conditions for the video diffusion model. However, the quality of these methods hinges on the accuracy of the predicted flow, which can easily falter without physical grounding, leading to unrealistic results. To overcome this limitation, we propose to take advantage of physics-based animation to generate physically consistent and plausible motions. This not only ensures geometric consistency across frames but also inherently respects fundamental physical laws, resulting in more realistic and reliable animations. 
\vspace{-4mm}
\paragraph{Geometry Registration} 
The first step to animate objects of interest in a given illustration is to establish a geometric basis. While previous works \cite{hong2023lrm, tang2025lgm} have explored predicting 3D geometry from single-view images, these methods are typically constrained to reconstructing real-world objects and do not generalize to in-the-wild anime images, since characters and objects in anime often lack an inherent 3D representation due to their stylized and flat nature. As shown in prior works on cartoon animation \cite{xing2016energy, kazi2016motion}, leveraging 2D animation techniques proves effective for generating anime-style dynamic effects. 
\begin{wrapfigure}{r}{0.55\linewidth}
\vspace{-0em}
\hspace{-1.6em}
    \includegraphics[width=1.1\linewidth]{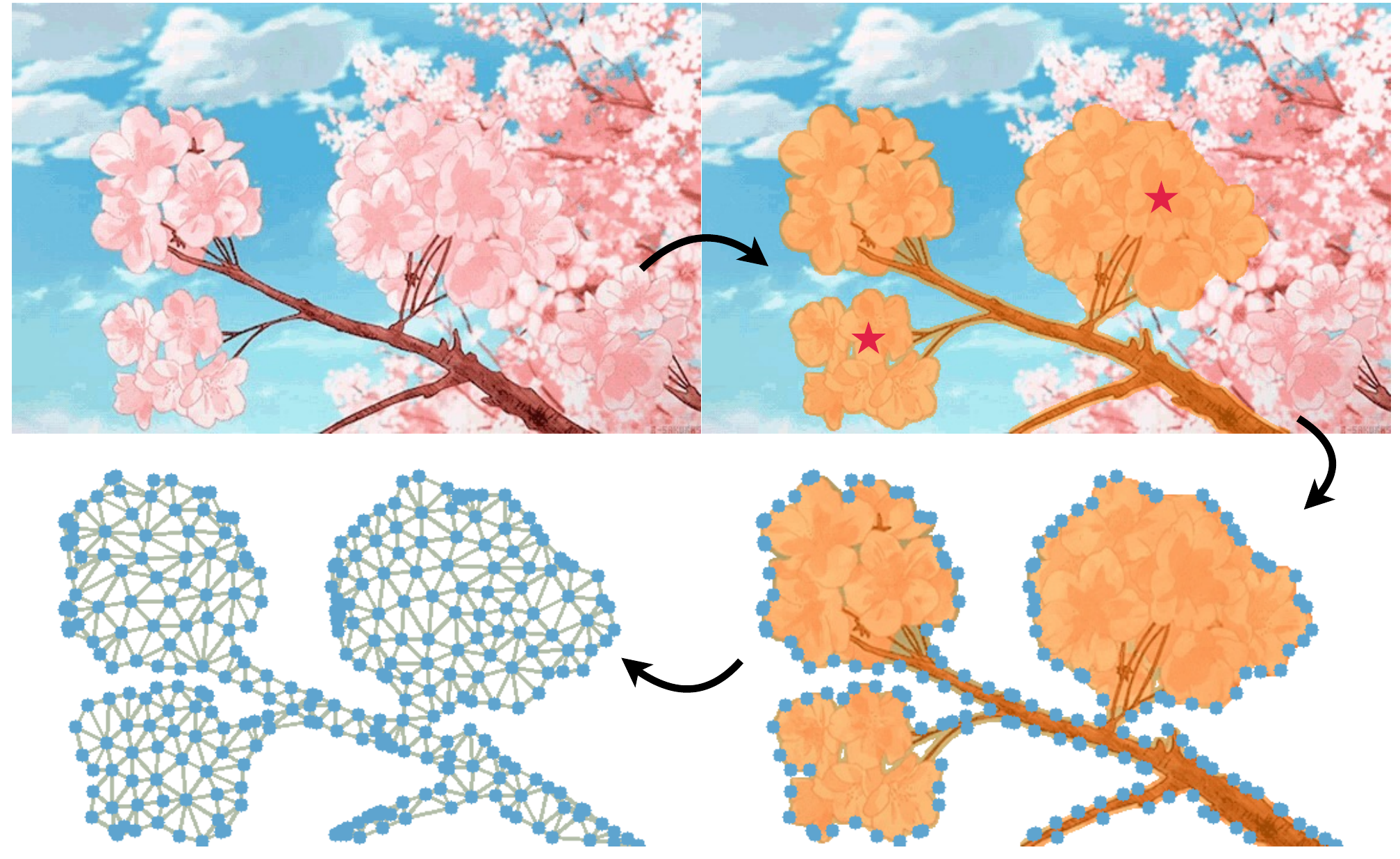}
  \vspace{-2em}
\end{wrapfigure}
\noindent Inspired by this, we focus on extracting 2D meshes for the objects of interest. To achieve this, we first utilize the Segment Anything Model (SAM) \cite{kirillov2023segment, ravi2024sam} to obtain segmentation masks for each target object, guided by user-specified query points. Along the contours of each segmentation mask, we uniformly sample boundary points, defining the outline of the intended mesh. We then employ conforming Delaunay triangulation \cite{lee1980two} with these boundary constraints to generate well-structured triangular meshes, creating a simulation-ready format for subsequent animation.
% \begin{figure}
%     \centering
%     \includegraphics[width=1.0\linewidth]{images/geometry_registration.pdf}
%     \caption{\textbf{Geometry Registration.} }
%     \label{fig:enter-label}
% \end{figure}
\vspace{-4mm}
\paragraph{Deformable Body Model}
Anime scenes often feature dynamic environmental effects such as external forces that interact with characters' clothing, hair, or other elements. Additionally, techniques like ``squash'' and ``stretch" are commonly used to convey motion and energy, adding expressiveness to animated objects. Motivated by these stylized dynamics, we model anime objects as deformable bodies, allowing them to capture the fluidity and exaggerated motion characteristic of anime. A key property of deformable bodies is their ability to change shape in response to external forces and attempt to return to their original rest shape upon deformation. To represent this physical behavior, we employ the Fixed Corotated constitutive model \cite{stomakhin2012energetically}, which defines the energy density as:
\begin{equation}
    \Psi(\bm{F}) = \mu \| \bm{F} - \bm{R} \|^2_F + \frac{\lambda}{2}(\text{det}(\bm{F}) - 1)^2,
\end{equation}
where $\mu$, $\lambda$ are the Lamé parameters, and $\bm{R}$ is the rotational part of $\bm{F}$ computed via polar decomposition. The first term models the stretching and compression resistance for the individual spatial directions and the second term describes resistance to volume change. The resulting internal forces $\bm{f}_{\text{int}}$ can then be derived via Eq.~(\ref{eq: internal_forces}). 
% To simulate the dynamic behavior of these deformable objects, we solve the motion equation (Eq.~\ref{eq: time_integration}) using a semi-implicit Euler method to compute the deformation map $\phi_i(X, t)$ at each time step $t$ for every triangle. Details of the resulting equations can be found in the supplementary materials \TODO{add this in supp}.
\vspace{-4mm}
\paragraph{Interactive Animation} While $\bm{f}_\text{int}$ governs the inherent physical behavior, $\bm{f}_\text{ext}$ allows for user-defined interactions. Inspired by the concept of energy strokes \cite{gilland2012elemental, xing2016energy}, we introduce customizable energy strokes that carry flow particles. These particles move along the user-specified strokes, propagating external forces to nearby vertices of the deformable mesh. This enables the creation of tailored animation effects, such as simulating wind interactions. 
Details of our flow particles can be found in the appendix. Additionally, we incorporate rigging point support, allowing animators to anchor specific regions or guide them along predefined trajectories, offering enhanced control and flexibility. Using the specified energy strokes and deformable modeling, we solve the motion equation (Eq.~(\ref{eq: time_integration})) to evolve the dynamics. We employ the semi-implicit Euler method to compute the deformation map $\phi_i(X, t)$ at each time step $t$ for every triangle. We refer to the appendix for detailed steps on solving the motion equation. For each pixel $\bm{X}_p$ in the reference image, if it lies within a triangle $\mathcal{T}_i$, we assign the displacement vector as $\bm{d}(\bm{p}) = \phi_i(\bm{X}_p, t) - \bm{X}_p$; otherwise, we assign a zero vector. Collecting displacement vector $\bm{d}(\bm{p})$ of all pixels yields the optical flow $\mathcal{F}_{0\rightarrow t}$, which defines the displacement fields for the reference illustration $I_0$ at time $t$.

\subsection{Generative Rendering}
In this section, we describe how to render the video sequence  $\{\hat{I}_1, \hat{I}_2, ...,\hat{I}_T\}$ given the reference image $I_0$ and the optical flow sequence $\{\mathcal{F}_{0 \rightarrow 1}, \mathcal{F}_{0 \rightarrow 2}, \ldots, \mathcal{F}_{0 \rightarrow T}\}$ derived from simulation.
\vspace{-4mm}
\paragraph{Sketch-Guided Rendering} While the video sequence can be directly generated by warping the reference image $I_0$ using optical flow fields, the resulting frames often exhibit black hole artifacts caused by occlusions. To address this problem, we extract the sketch of $I_0$, denoted as $S_0$, and obtain the sketch at time $t$ as follows: 
\begin{equation} S_t = \mathcal{W}(S_0, \mathcal{F}_{0 \rightarrow t}, w_{0 \rightarrow t}), \end{equation} 
where $\mathcal{W}$ represents the forward-warping operator, and $w_{0 \rightarrow t}$ denotes the warping weights for each pixel in $I_0$. In forward warping, multiple pixels may map to the same 2D location in the output frame \cite{niklaus2020softmax}, potentially leading to artifacts or distortions if the weights $w_{0 \rightarrow t}$ are not properly defined. Inspired by \cite{li2024generative}, we set the pixel weight as $w_{0 \rightarrow t}(\bm{p}) = \|\mathcal{F}_{0 \rightarrow t}(\bm{p})\|_2$, giving higher importance to pixels with larger motion, typically corresponding to foreground objects.

Next, we leverage a video diffusion model to generate the rendered frames $\{ I^r_1, I^r_2, ..., I^r_T\}$ using the obtained sketch sequence $\{S_1, S_2, ..., S_T\}$. The diffusion model uses the reference image $I_0$ as an input, and we employ a ControlNet \cite{zhang2023adding} with the sketches as control signals, guiding the generation process to ensure that the results align with the sketch inputs. We observed that during the inference time, due to segmentation inaccuracies, the optical flow warping may introduce unintended distortions, or miss parts of the object contour, generating imperfect sketches. To address this, we apply Gaussian blur to the input sketches at both training and inference time, which smooths out inconsistencies. The video diffusion model is then capable of refining the results, leveraging its generative capabilities to handle imperfections and produce coherent outputs.
\vspace{-4mm}
\paragraph{Complementary Dynamics}
Unlike motions in the real world, dynamic effects in animation do not strictly adhere to physical laws and can not be fully captured by 2D animation methods. In the industrial animation pipeline, artists typically begin by creating a series of keyframes that define the primary motion trajectory, followed by drawing in-between frames to ensure smooth and fluid transitions. Inspired by this workflow, we leverage a data-driven module to enhance the physics-based animated results. Specifically, we select keyframes from the sketch-guided rendering results, forming the keyframe sequence $\{I^r_0, I^r_{n}, I^r_{2n}, ..., I^r_{in}\}$, where $n$ denotes the gap between keyframes. We then employ a cartoon interpolation video diffusion model \cite{xing2024tooncrafter}, which synthesizes intermediate frames $\{\hat{I}_{in+1}, \hat{I}_{in+2}, ..., \hat{I}_{i(n+1)-1}\}$ between each adjacent keyframe pair $(I^r_{in}, I^r_{i(n+1)})$. For keyframes, we simply assign $\hat{I}_{in} =I^r_{in}$. This approach allows us to introduce expressive, data-driven complementary dynamics that go beyond what can be achieved through physics-based animation alone.

\label{sec:method}
\section{Experiments}
In this section, we conduct a comprehensive comparison of our method against existing video diffusion models and demonstrate that our approach generates high-quality and physically plausible animations.

\vspace{-15px}
\paragraph{Implementation Details} We implement our deformable body simulator using Taichi \cite{hu2019taichi}. During interactive animation, users can adjust the Lamé parameters $\mu$ and $\lambda$ to control the characteristics of the objects based on their specific needs. For generative rendering, we follow LVCD \cite{huang2024lvcd} to train a sketch-guided ControlNet for the stable video diffusion model \cite{chai2023stablevideo}, using blurred sketches as control signals to address potential imperfections in the predicted sketch sequence. When utilizing ToonCrafter \cite{xing2024tooncrafter} for generating in-between frames, similar to its original implementation, we also train an additional sketch-guided ControlNet, but set the control scale to 0.1 during inference. We also set $n=15$ for the in-betweening step. This configuration helps the generated frames follow the rough motion indicated by the sketches while introducing extra stylized animation details.
\vspace{-15px}
\paragraph{Dataset}
We build our training dataset using the Sakuga-42M Dataset \cite{pan2024sakuga}, which consists of 1.4 million animation video clips. To ensure high-quality training samples, we filter out clips with fewer than 24 frames and dynamics scores outside the range of 0.05 to 0.7. Next, we extract sketches from the selected clips using the method from \cite{chan2022drawings}, resulting in a final dataset of 380,000 pairs of sketches and corresponding video sequences.
\vspace{-15px}
\paragraph{Baseline}
We compare our method against two categories of video generation approaches. The first category includes state-of-the-art Image-to-Video (I2V) models, such as Cinemo \cite{ma2024cinemo} and DynamiCrafter \cite{xing2025dynamicrafter}, which take an input image and use text prompts to guide the motion dynamics of the generated videos. For these models, we generate prompts describing the image content and expected motion using ChatGPT-4V. The second category, including Motion-I2V \cite{shi2024motion} and Drag Anything \cite{wu2025draganything}, usually trains an additional motion-control module for video generative models such that, given a single image input, a user-specified trajectory can be provided to control the movement of objects in the generated frames. To ensure a fair comparison, we extract trajectories from our animated results and use them as input for these methods.

% \subsection{Results}
% We show \TODO{a gallery of figure}

% \subsection{Comparison Study}
% \paragraph{Evaluation Metrics

\begin{figure*}
    \vspace{-10px}
    \centering
    \includegraphics[width=1.0\linewidth]{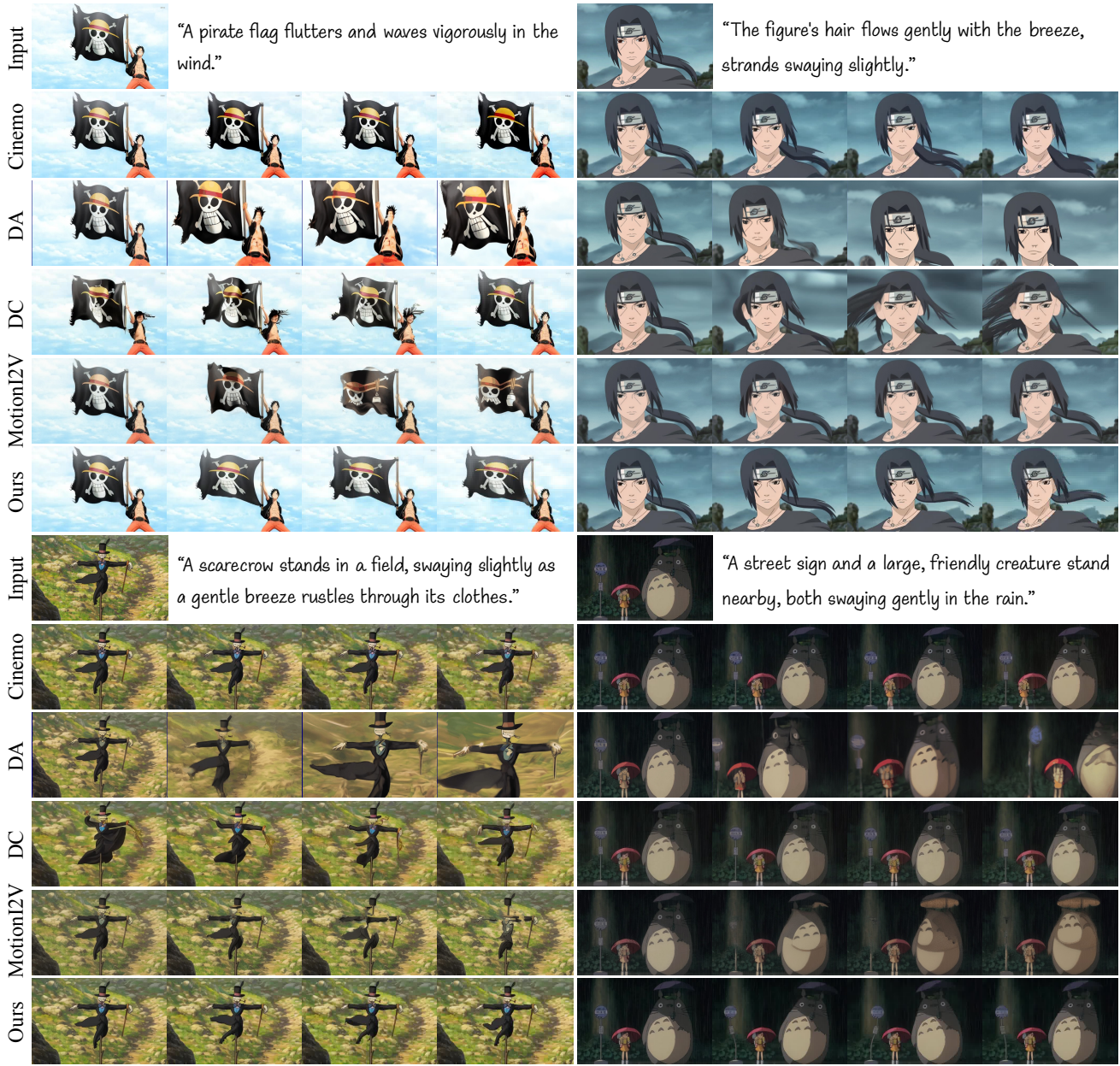}
    \vspace{-20px}
    \caption{\textbf{Qualitative Comparison}. We compare our results against Cinemo \cite{ma2024cinemo}, Drag Anything \cite{wu2025draganything}, DynamiCrafter \cite{xing2025dynamicrafter} and Motion-I2V \cite{shi2024motion}. Text prompts for Cinemo and DynamiCrafter are generated using ChatGPT-4V, while trajectories for Drag Anything and Motion-I2V are extracted from our animated results.}
    \label{fig:qualitative_comparison}
    \vspace{-15px}
\end{figure*}

\subsection{Quantitative Evaluation}
Given the absence of established benchmarks for anime-style image-to-video generation, we follow Motion-I2V \cite{shi2024motion} and construct a test set comprising 20 anime images with stylized dynamic elements such as swaying hair, clothing, and plants. For each method, we generate 10 video samples per image, resulting in a total of 200 videos per method. To quantitatively evaluate the generated results, we use the Fréchet Inception Distance (FID) \cite{heusel2017gans} to measure the similarity between the generated frames and the reference images. Additionally, we employ VideoScore \cite{he2024videoscore} to assess multiple aspects of video quality, including visual quality, temporal consistency, dynamic degree, and factual consistency. The factual consistency evaluates the consistency of the video content with common sense and factual knowledge. We exclude the text alignment score as our method and the trajectory-controlled approach do not involve text input.

The quantitative evaluation results, presented in Tab.~\ref{tab:quantitative comparison}, demonstrate that our method outperforms baseline approaches across most metrics, including visual quality, temporal consistency, and factual consistency, demonstrating the overall high quality of our generated videos. As shown in Fig.~\ref{fig:qualitative_comparison}, Cinema tends to generate static videos that closely resemble the reference image, leading to better FID scores. While the dynamic degree score for DragAnything is notably higher than other methods, this is due to its tendency to misinterpret motion control as camera movement, causing shifts in the entire image space and resulting in an inflated dynamics score. Although our method's dynamic degree score is slightly lower than that of DynamiCrafter and Motion-I2V, these methods often exhibit implausible, distorted motions that create an illusion of increased dynamics. In contrast, our approach produces geometry-consistent motions and achieves high factual consistency, demonstrating the realistic motion dynamics enabled by our physics-based modeling.
In addition, following \cite{kolkin2019style, tokenflow2023}, we conducted an user study adopting a two-alternative forced choice (2AFC) protocol, where participants are asked to choose the preferred video based on temporal consistency, visual quality, motion plausibility, and overall feeling given two videos (one from our method and one from baselines). The user preference results, presented in Tab.~\ref{tab:user study}, show that our proposed method consistently outperforms the baselines across all evaluation criteria.

\subsection{Qualitative Comparison}
\vspace{-10px}
We present a qualitative comparison with baseline methods in Fig.~\ref{fig:qualitative_comparison}. Cinemo often produces static results with minimal motion dynamics. Drag Anything frequently misinterprets the motion trajectory as a camera movement, resulting in unintended dynamic sequences. DynamiCrafter, while generating larger motions, struggles to maintain the geometry of input objects, leading to noticeable distortions. Motion-I2V has difficulty preserving the appearance of the input content, often yielding unsatisfactory results. In contrast, our physics-guided approach ensures both physically plausible motions and high-quality rendering, preserving the geometry and visual consistency of the input.

\begin{table}[t]
\centering
\caption{\textbf{Quantitative Comparisons.} We report FID to evaluate the similarity between the reference image and generated frames. VSVQ, VSTC, VSDD, and VSFC represent scores for visual quality, temporal consistency, dynamic degree, and factual consistency, respectively, as measured by VideoScore \cite{he2024videoscore}. \vspace{-5px}} 
\setlength\tabcolsep{1.8pt}
\label{tab:quantitative comparison}
\small{
\begin{tabular}{p{1.0in}cccccc} 
\hline
\textbf{Methods} & FID$\downarrow$ & VSVQ$\uparrow$ & VSTC$\uparrow$  & VSDD$\uparrow$ & VSFC$\uparrow$ \\ \hline  %physics score, text score, dynamic score, user consistency score, user preference score
Cinemo \cite{ma2024cinemo}& \cellcolor{Apricot}{49.5} & 2.85 & 2.80 & 2.42 & 2.58 \\ 
DragAnything \cite{wu2025draganything}& 148.9 & 2.77 & 2.45 & \cellcolor{Apricot}{2.97} & 2.52 \\ 
DynamiCrafter \cite{xing2025dynamicrafter}& 94.9 & 2.78 & 2.68 & 2.53 & 2.51 \\ 
Motion-I2V \cite{shi2024motion}& 121.8 & 2.70 & 2.50 & 2.66 & 2.39 \\ 
\hline
Ours & 90.4 & \cellcolor{Apricot}{2.89} & \cellcolor{Apricot}{2.86} & 2.48 & \cellcolor{Apricot}{2.64}\\ 
\hline
\end{tabular}
\vspace{-5px}
}
\end{table}

\begin{table}[t]
\centering
\caption{\textbf{User Study.} We show the user preference for our method over the baseline methods in terms of visual quality (VQ), temporal consistency (TC), motion plausibility (MP), and overall feeling. \vspace{-10px}} 
\setlength\tabcolsep{6.0pt}
\label{tab:user study}
\small{
\begin{tabular}{p{1.0in}cccccc} 
\hline
\textbf{Methods} & VQ & TC & MP  & Overall \\ \hline  %physics score, text score, dynamic score, user consistency score, user preference score
Cinemo \cite{ma2024cinemo}& 86\% & 83\% & 82\% & 81\% \\ 
DragAnything \cite{wu2025draganything}& 93\% & 91\% & 89\% & 91\% \\ 
DynamiCrafter \cite{xing2025dynamicrafter}& 84\% & 78\% & 76\% & 81\% \\ 
Motion-I2V \cite{shi2024motion}& 95\% & 94\% & 97\% & 96\% \\ 
\hline
\end{tabular}
}
\end{table}

% Furthermore, current motion-guided video diffusion methods primarily control the intended movement by offering users two main options: either specifying a drag trajectory for the desired motion path or indicating a general motion direction and adjusting the strength to control its magnitude. Nevertheless, these methods cannot often capture complex object deformations, as they do not consider the underlying material properties. In contrast, our physics-based approach directly models the physical behaviors of the objects, incorporating material characteristics such as elasticity and stiffness. By leveraging material parameters, our method naturally allows users to intuitively adjust the object's response under external forces. \TODO{add one figure for this}.

\subsection{Additional Qualitative Results}
\begin{figure}
    \centering
    \includegraphics[width=1.0\linewidth]{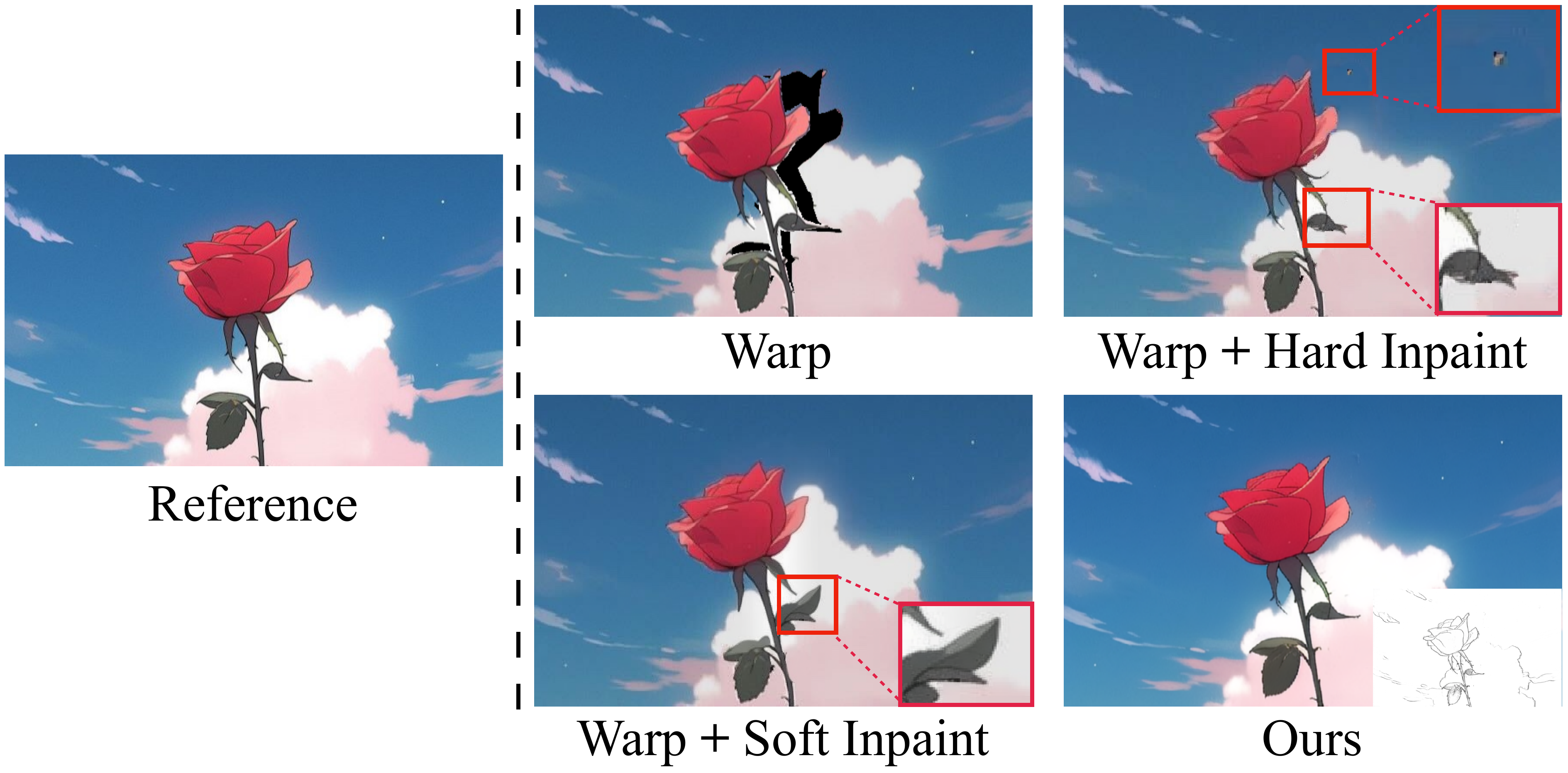}
    \caption{\textbf{Sketch-Guided Rendering}. Applying warping and inpainting introduces artifacts due to segmentation inaccuracy. Soft-inpainting \cite{levin2023differential} reduces these issues but can alter the content. Our sketch-guided rendering method produces high-quality results while preserving image details.}
    \label{fig:rendering_ablation}
    \vspace{-15px}
\end{figure}
\paragraph{Sketch-Guided Rendering}
After physics-based animation in our proposed framework, we obtain a sequence of optical flow $\{\mathcal{F}_{0 \rightarrow 1}, \mathcal{F}_{0 \rightarrow 2}, \ldots, \mathcal{F}_{0 \rightarrow T}\}$. A straightforward approach would be to warp the input image $I_0$ using these optical flows and apply an inpainting algorithm to fill any occluded areas. However, this often results in suboptimal outputs. The main issue lies in segmentation inaccuracies, which can cause unintended regions to be warped or leave parts of the intended objects static, as shown in Fig.~\ref{fig:rendering_ablation}. While introducing a soft mask \cite{levin2023differential} into the inpainting process can help smooth the transition between the newly generated region and the original figure, it may also alter the content, making it diverge from the appearance of the reference image. Additionally, applying inpainting frame by frame may also cause temporal consistency issues. To address this, our method employs a sketch-guided rendering module. We first extract a sparse geometric representation of the image and apply the animation dynamics directly to this sketch-based structure. This sparse representation is more robust to segmentation inaccuracies and helps preserve the intended motion. Our subsequent rendering module then synthesizes high-quality video sequences from the animated sketches, maintaining both temporal consistency and visual fidelity.

% \TODO{maybe use the rose example to show we can not directly warp the original frame}

% Here we show the necessity of

\begin{figure}
    \centering
    \includegraphics[width=1.0\linewidth]{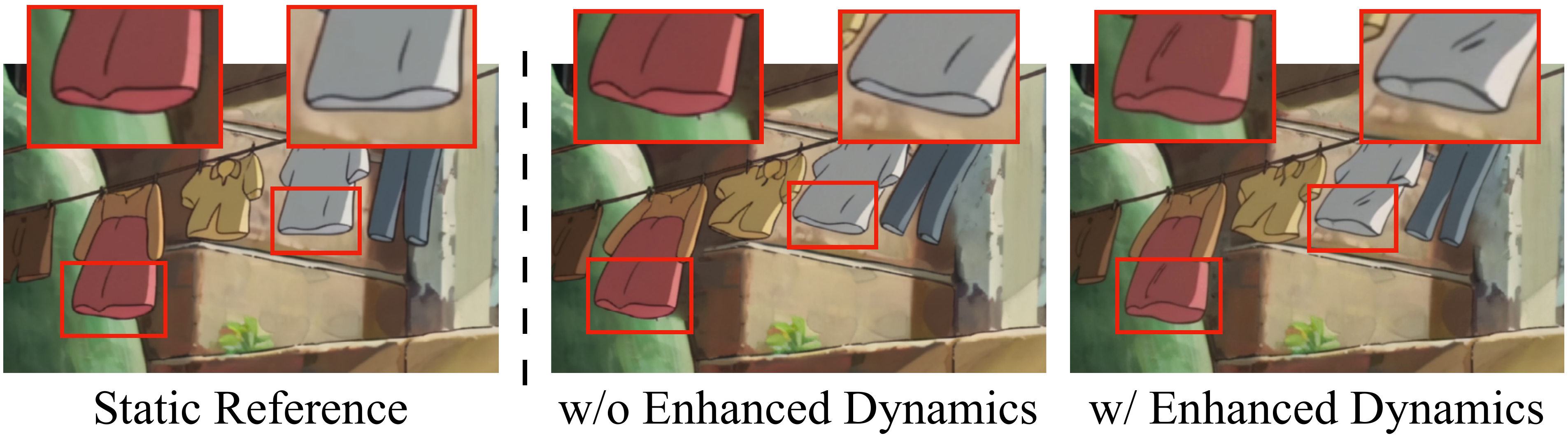}
    \vspace{-20px}
    \caption{\textbf{Complementary Dynamics Enhancement}. While physics-based animation maintains geometric consistency, it may lack the fluidity and exaggeration commonly seen in anime. We employ a data-driven interpolation module to enhance the motion dynamics, creating more natural-looking animations that better resemble real anime.}
    \label{fig:complementary_dynamics_ablation}
    \vspace{-10px}
\end{figure}
\vspace{-15px}
\paragraph{Complementary Dynamics Enhancement}
While our physics-based animation ensures physically accurate and geometry-consistent motion, it is inherently limited to 2D representations and cannot fully capture 3D effects. Moreover, the exaggerated dynamics often seen in anime do not always conform to strict physical laws. To address these limitations, we incorporate a data-driven cartoon interpolation module \cite{xing2024tooncrafter} to introduce complementary dynamics. As illustrated in Fig.~\ref{fig:complementary_dynamics_ablation}, this interpolation module enables fluid contour deformations of clothing during motion, rather than rigidly adhering to static geometry. It also dynamically generates new sketch lines, enhancing the expressiveness of the animation and bringing it closer to the aesthetic of traditional hand-drawn anime. 
% To further evaluate the effectiveness of the interpolation module, we conduct a user study comparing animations before and after applying the dynamics enhancement. Participants are asked, ``Which video resembles real anime more?'' The results show a significant preference (73.8\%) for the enhanced animations, demonstrating the module’s contribution to improved visual quality and stylization.

% \begin{table}[t]
% \centering
% \caption{\textbf{Ablation Study.} Our approach \TODO{fill the score} . \vspace{-5px}} 
% \setlength\tabcolsep{1.8pt}
% \label{tab: ablation}
% \small{
% \begin{tabular}{p{0.75in}cccc} 
% \hline
% \textbf{Methods} & \textbf{PS} & \textbf{TS} & \textbf{DS} \\ \hline
% without x1 & 0.0000 & \cellcolor{Goldenrod}{0.0000} & \cellcolor{Goldenrod}{0.0000} \\ 
% without x2 & 0.0000 & 0.0000 & 0.0000 \\ 
% Whole & \cellcolor{Goldenrod}{0.0000} & 0.0000 & 0.0000 \\ 
% \hline
% Ours & \cellcolor{Apricot}{0.0000} & \cellcolor{Apricot}{0.0000} & \cellcolor{Apricot}{0.0000} \\ 
% \hline
% \end{tabular}
% \vspace{-10px}
% }
% \end{table}

% \section{Limitation and Future Work}
% \label{sec:limitation}
\section{Conclusion}
\label{sec:conclusion}
% \paragraph{Limitation}
% \TODO{}

% \paragraph{Conclusion}
We present \METHOD, a novel framework for generating dynamic and stylized animations from static anime illustrations by integrating physics-based simulations with data-driven generative models. Our approach generates physically plausible, fluid and exaggerated anime-styled motion through deformable body simulations, providing controllability to users through user-guided energy strokes. A sketch-guided video diffusion model ensures high-quality, temporally consistent frames, while a data-driven anime frame interpolation model adds expressive, non-physical dynamics. The experiments show that our proposed method outperforms existing video diffusion methods, offering a powerful tool for creating visually compelling and user-controllable anime-style animations. 
% This work bridges the gap between traditional hand-drawn techniques and modern generative models, paving the way for future extensions to a broader range of artistic styles and effects.
\section*{Acknowledgements}
We acknowledge support from NSF 2153851. We would like to thank Rahul Garg, Hossein Taghavi, Roshni Cooper, Ritwik Kumar, Boris Chen, Amli Murphy, Oliver Banasiak and Taiki Sakurai for their valuable suggestions and support.

{
    \small
    \bibliographystyle{ieeenat_fullname}
    \bibliography{main}
}

\clearpage
\newpage
\appendix
\section*{\Large Appendix}
\section{Motion Equation}
To model the fluidity and exaggerated motions typical of anime, we represent anime objects as deformable bodies and adopt the Fixed Corotated model \cite{stomakhin2012energetically} as the constitutive model. The dynamics of the deformable bodies are governed by Newton's Second Law, expressed as:
\begin{equation}\label{eq: motion equation}
    \frac{d^2 \bm{x}}{dt^2} = \bm{M}^{-1}(\bm{f}_{\text{int}}(\bm{x}) + \bm{f}_{\text{ext}}(\bm{x})),
\end{equation}
where $\bm{M}$ is the mass matrix, representing the masses of all vertices:
\begin{equation}
    \bm{M} = \left(\begin{array}{cccc}
                m_1 & 0 & \cdots & 0 \\
                0  & m_2 & \cdots & 0 \\
                \vdots & \vdots & \ddots & \vdots \\
                0& 0& 0& m_N
                \end{array}\right).
\end{equation}
The mass of $i$-th vertex is calculated as
\begin{equation}
    m_i = \sum\limits_{\mathcal{T}_j \in S_i} \frac{\rho_j V_j}{3}
\end{equation}
where $S_i$ is the set of triangles containing the $i$-th vertices. $\rho_j$ and $V_j$ denote the mass density and volume of $j$-th triangle, respectively.
The internal force fir the $\bm{f}_{\text{int}}(\bm{x})$ of Fixed Corotated deformable model can be derived as
\begin{equation}
\begin{aligned}
\bm{f}_{\text{int}}(\bm{x}) & = -\frac{\partial E(\bm{x})}{\partial \bm{x}} \\
    & = - \frac{\partial\sum\limits_{i=1}^N \Psi(\bm{F}_i) V_i}{\partial \bm{x}} \\
    & = - \sum\limits_{i=1}^N \left( \frac{\partial \Psi(\bm{F}_i)}{\partial\bm{F}_i} \frac{\partial \bm{F}_i}{\partial \bm{x}} V_i\right), \\
\end{aligned}
\end{equation}
where 
\begin{equation}
    \begin{aligned}
        \frac{\partial \Psi(\bm{F}_i)}{\partial\bm{F}_i} &= \frac{\partial(\mu \| \bm{F}_i - \bm{R}_i \|^2_F + \frac{\lambda}{2}(\text{det}(\bm{F}_i) - 1)^2)}{\partial\bm{F}_i} \\
        &= 2\mu (\bm{F}_i - \bm{R}) + \lambda \left(\text{det}(\bm{F}_i) - 1)\right) \text{det}(\bm{F}_i) \bm{F}_i^{-T},
    \end{aligned}
\end{equation}
The deformation gradient $\bm{F}_i$ for a 2D triangle $\mathcal{T}_i$ is computed as:
\begin{equation}
    \begin{aligned}
        \bm{F}_i =  [\bm{x}_1 - \bm{x}_0, \bm{x}_2 - \bm{x}_0] [\bm{X}_1 - \bm{X}_0, \bm{X}_2 - \bm{X}_0]^{-1}.
    \end{aligned}
\end{equation}
Here $\{\bm{X}_0, \bm{X}_1, \bm{X}_2\}$ and $\{\bm{x}_0, \bm{x}_1, \bm{x}_2\}$ represent the positions of vertices of triangle $\mathcal{T}_i$ in the undeformed (material) space and the deformed (world) space respectively.

To solve the motion equation \eqref{eq: motion equation}, we apply a semi-implicit Euler method, discretizing it as:
\begin{equation}\label{eq: semi-implicit euler}
    \begin{aligned}
         \bm{x}^{n+1} &= \bm{x}^n + \Delta t \bm{v}^{n+1}, \\
         \bm{v}^{n+1} &= \bm{v}^n + \Delta t \bm{M}^{-1} (\bm{f}_{\text{int}}(\bm{x}^n) + \bm{f}_{\text{ext}}(\bm{x}^n)),
    \end{aligned}
\end{equation}
where $\bm{x}^n, \bm{v}^n$ and $\bm{x}^{n+1}, \bm{v}^{n+1}$ denote the positions and velocities at time steps $t^n$ and $t^{n+1}$, respectively. The timestep size $\Delta t$ is set to 0.001 in our experiments. By iteratively applying Eq.~\ref{eq: semi-implicit euler} from $t = 0$, we obtain the physical states (positions and velocities) at any future time step.

\section{Details of Interactive Animation}
\begin{figure}
    \centering
    \includegraphics[width=1.0\linewidth]{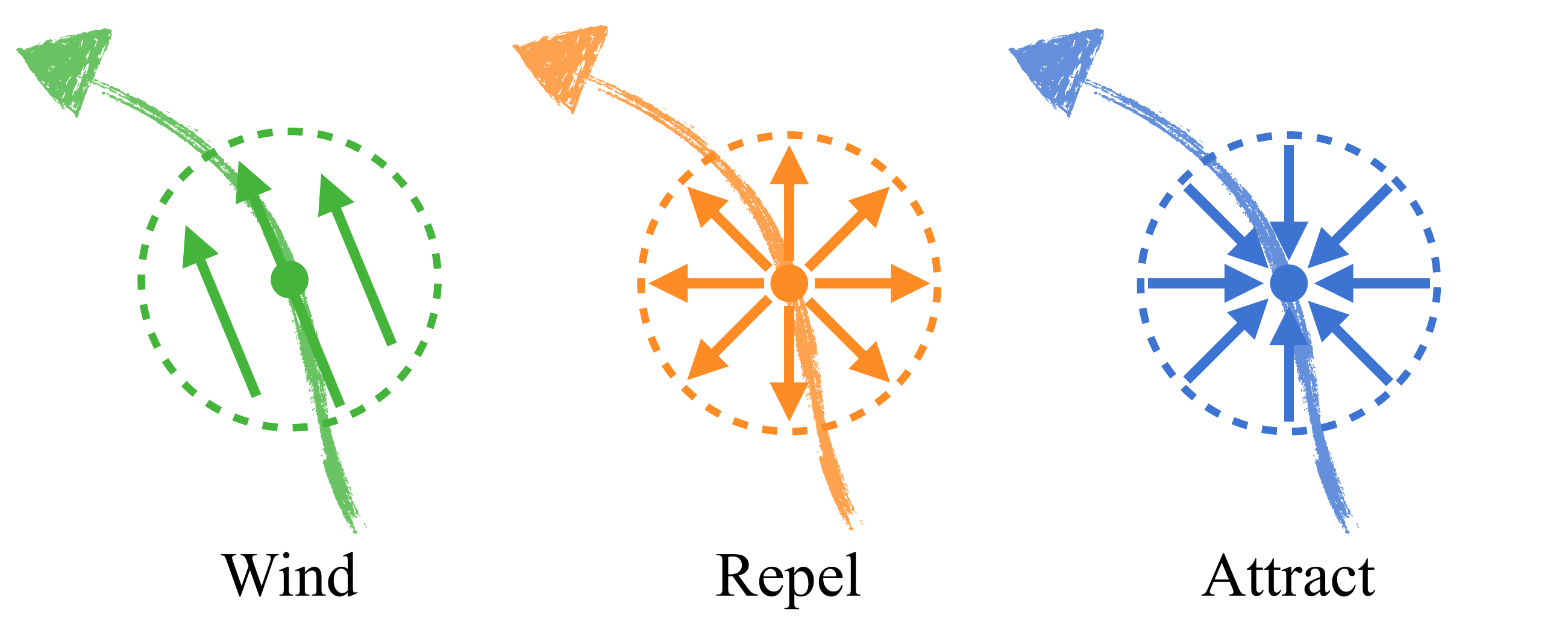}
    \caption{\textbf{Flow Particles}. Wind particles can simulate wind effects, creating natural swaying motions; repel particles act as collision barriers, adjusting object trajectories; attract particles can pull objects toward specific points as desired.}
    \label{fig:flow_particles}
    \vspace{-10px}
\end{figure}
To enable interactive animation, we introduce energy strokes that carry flow particles, which exert forces on nearby vertices to drive motion. As illustrated in Fig.~\ref{fig:flow_particles}, we provide three types of customizable energy strokes: wind, repel, and attract.

\begin{figure*}
    \centering
    \includegraphics[width=1.0\linewidth]{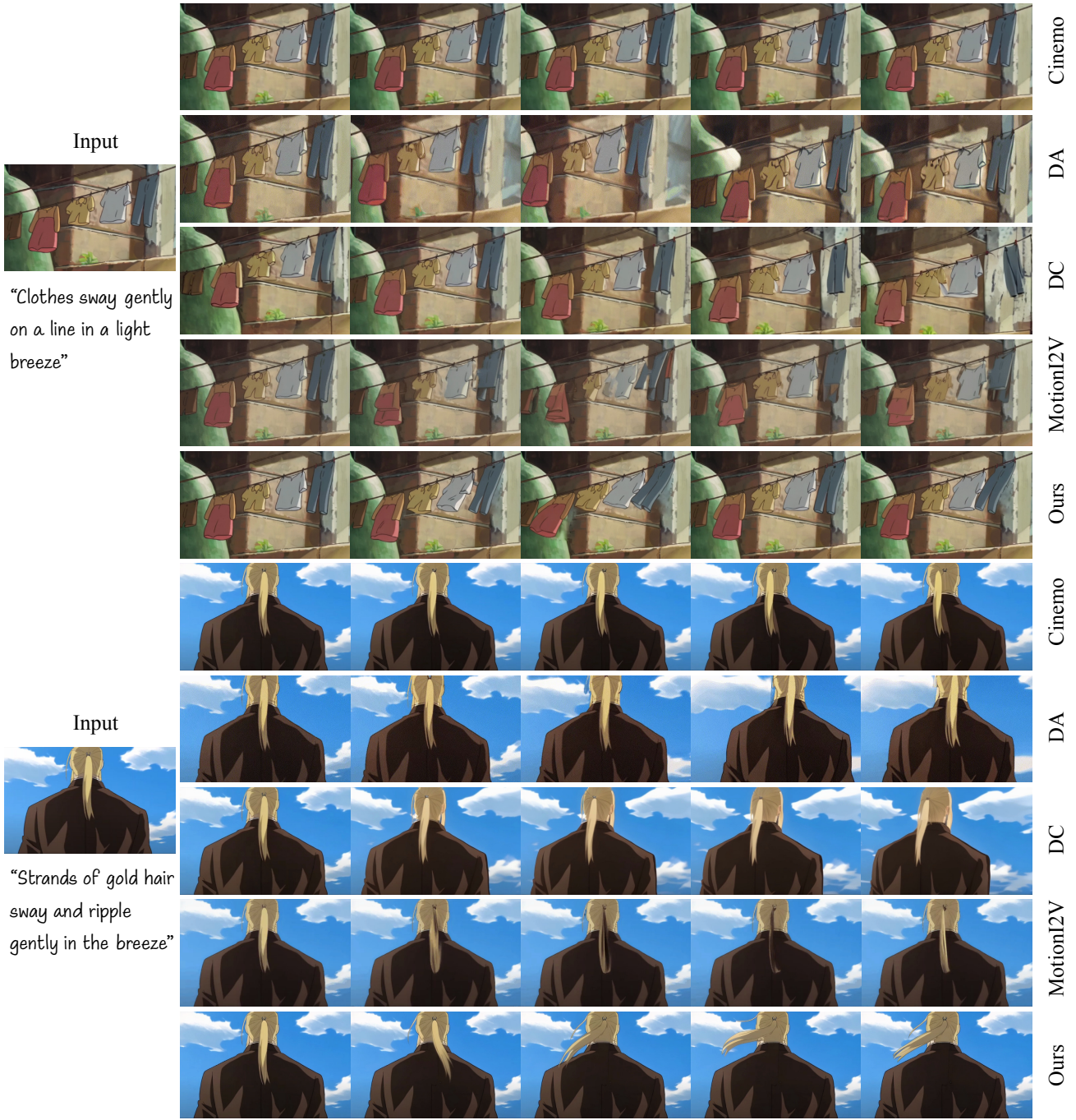}
    \caption{\textbf{Additional Qualitative Comparison}. We provide additional comparison results against Cinemo \cite{ma2024cinemo}, Drag Anything (DA) \cite{wu2025draganything}, DynamiCrafter (DC) \cite{xing2025dynamicrafter} and Motion-I2V \cite{shi2024motion}.}
    \label{fig:sup_qua1}
\end{figure*}

\begin{figure*}
    \centering
    \includegraphics[width=1.0\linewidth]{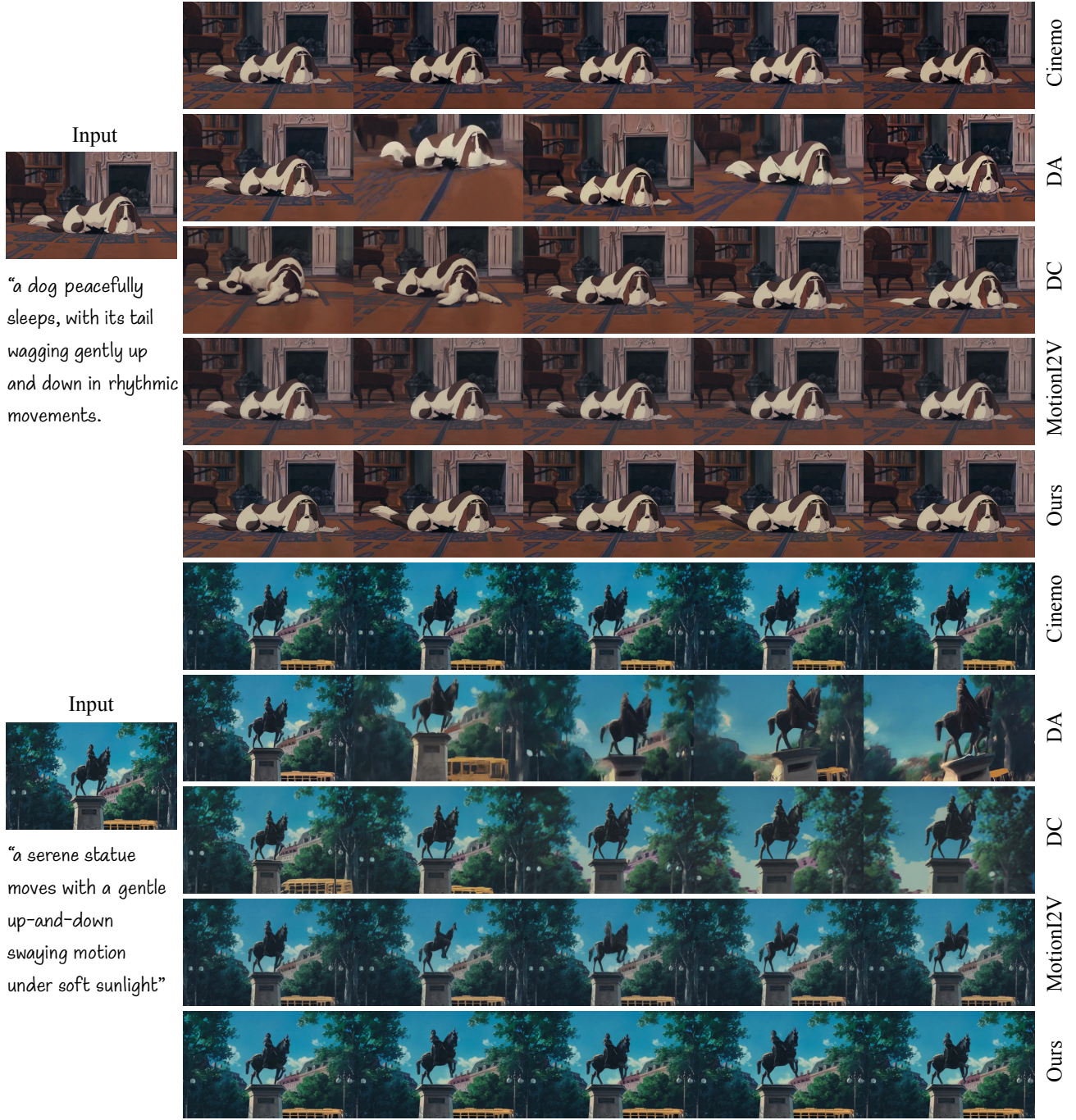}
    \caption{\textbf{Additional Qualitative Comparison}.  We provide additional comparison results against Cinemo \cite{ma2024cinemo}, Drag Anything (DA) \cite{wu2025draganything}, DynamiCrafter (DC) \cite{xing2025dynamicrafter} and Motion-I2V \cite{shi2024motion}.}
    \label{fig:sup_qua2}
\end{figure*}

\begin{figure*}
    \vspace{-10px}
    \centering
    \includegraphics[width=1.0\linewidth]{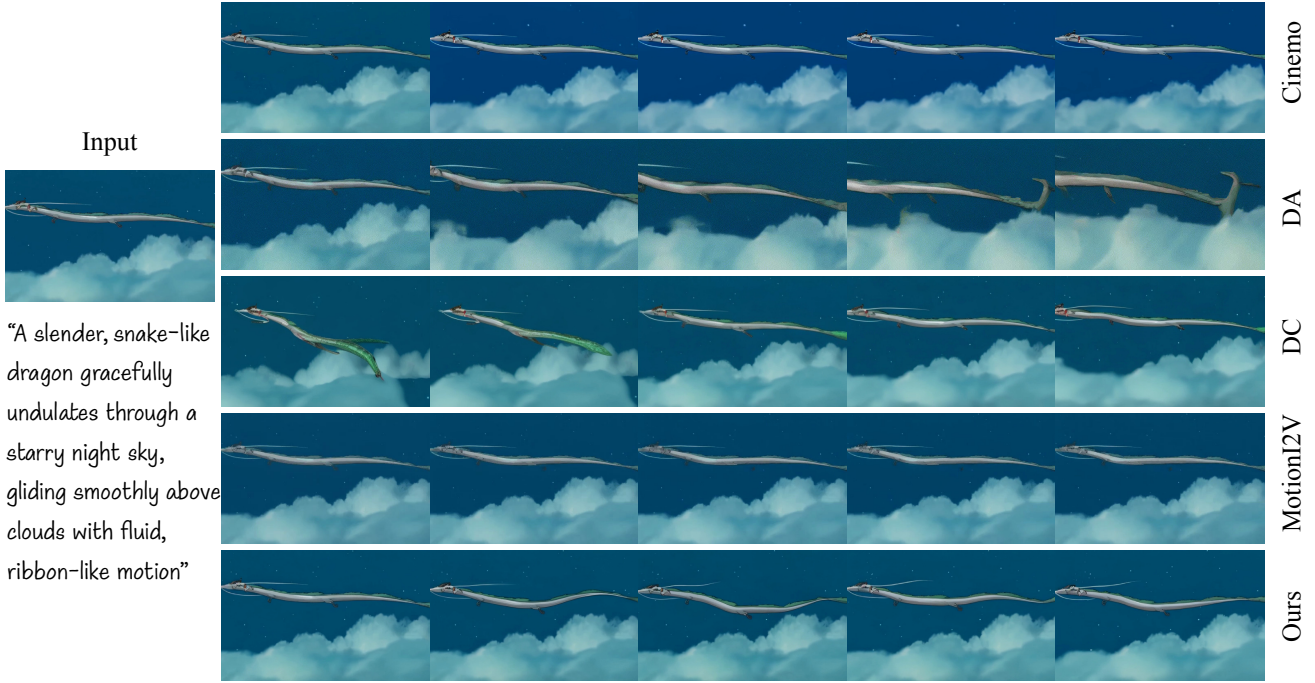}
    \caption{\textbf{Additional Qualitative Comparison}.  We provide additional comparison results against Cinemo \cite{ma2024cinemo}, Drag Anything (DA) \cite{wu2025draganything}, DynamiCrafter (DC) \cite{xing2025dynamicrafter} and Motion-I2V \cite{shi2024motion}.}
    \label{fig:sup_qua3}
    \vspace{-10px}
\end{figure*}

For a wind flow particle located at position $\bm{p}$ and a nearby vertex at position $\bm{q}$, the exerted force is defined as:
\begin{equation}
    \begin{aligned}
        \bm{f}_{\text{wind}}(\bm{p}, \bm{q}) &= s(1 - w(\bm{p}, \bm{q})) \bm{d},\\
        w(\bm{p}, \bm{q}) &= \frac{\|\bm{p} - \bm{q}\|}{r},
    \end{aligned}
\end{equation}
where $r$ is the influence range, $s$ represents the particle's strength, and $\bm{d}$ denotes the particle's movement direction. Similarly, the forces exerted by repel and attract flow particles are given as:
\begin{equation}
    \begin{aligned}
        \bm{f}_{\text{repel}} &= s\frac{\bm{q} - \bm{p}}{r},\\
        \bm{f}_{\text{attract}} &= s\left(1 - \frac{\|\bm{q} - \bm{p}\|}{r}\right) \frac{\bm{p} - \bm{q}}{\|\bm{p} - \bm{q}\|}.
    \end{aligned}
\end{equation}

In addition to flow particles, we incorporate rigging point support to allow animators to fix specific regions or define desired motion trajectories. Inspired by the swaying effects commonly seen in anime, we design a periodic wavy rigging point. The position of a wavy rigging point at time $t$ is expressed as:
\begin{equation}
    \bm{x}_r(t) = \bm{X}_r + s\sin(ft) \bm{d}_r,
\end{equation}
where $\bm{X}_r$ is the position in the undeformed state, $f$ is the swaying frequency, and $\bm{d}_r$ represents the swaying direction. The selected region will follow the rigging point's motion, creating desired animation effects.

\section{Training Details}
We adopt the standard ControlNet \cite{zhang2023adding} design, integrating each frame control signal into the ToonCrafter \cite{xing2024tooncrafter}. For LVCD \cite{huang2024lvcd}, we use the same network architecture but preprocess the input sketch with Gaussian blur. Both are trained on 8 A100 GPUs with the learning rate 1e-5 and batch size 32. The ControlNet in ToonCrafter is trained from the SD2.1 model for 50K steps. LVCD is fine-tuned from its pretrained for 10K steps. We will carefully add these details to our revised appendix.

\section{More Results}
\subsection{Controllable Generation}
\begin{table}[t]
\centering
\caption{\textbf{Ablation of Dynamics Enhancement.} We present the evaluation results for our Dynamics Enhancement (DE) module. The metrics VSVQ, VSTC, VSDD, and VSFC represent VideoScore \cite{he2024videoscore} assessments of visual quality, temporal consistency, dynamic degree, and factual consistency, respectively. User preferences are determined based on which video resembles real anime more.
% We report FID to evaluate the similarity between the reference image and generated frames. VSVQ, VSTC, VSDD, and VSFC represent scores for visual quality, temporal consistency, dynamic degree, and factual consistency, respectively, as measured by VideoScore \cite{he2024videoscore}.
} 
\setlength\tabcolsep{1.8pt}
\label{tab:ablation_study}
\small{
\begin{tabular}{p{0.75in}cccccc} 
\hline
\textbf{Methods} & User Pref.$\uparrow$ & VSVQ$\uparrow$ & VSTC$\uparrow$  & VSDD$\uparrow$ & VSFC$\uparrow$ \\ \hline  %physics score, text score, dynamic score, user consistency score, user preference score
Ours w/o DE& 29.6\% & 2.94 & 2.90 & 2.48 & 2.68 \\ 
Ours w/ DE & 70.4\%  & 2.89 & 2.86 & 2.48 & 2.64 \\ 
\hline
\end{tabular}
}
\end{table}

\begin{figure}
    \centering
    \includegraphics[width=1.0\linewidth]{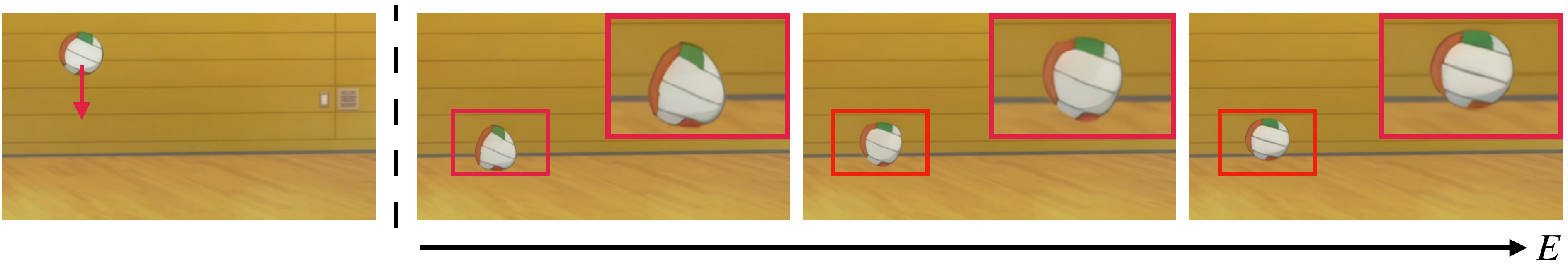}
    \caption{\textbf{Controllable Generation with Material Property.} As the Young's modulus $E$ of the ball increases from left to right, the ball becomes more capable of maintaining its original shape under external forces.}
    \label{fig:controllabel_generation}
    \vspace{-10px}
\end{figure}

\begin{figure}
    \centering
    \includegraphics[width=1.0\linewidth]{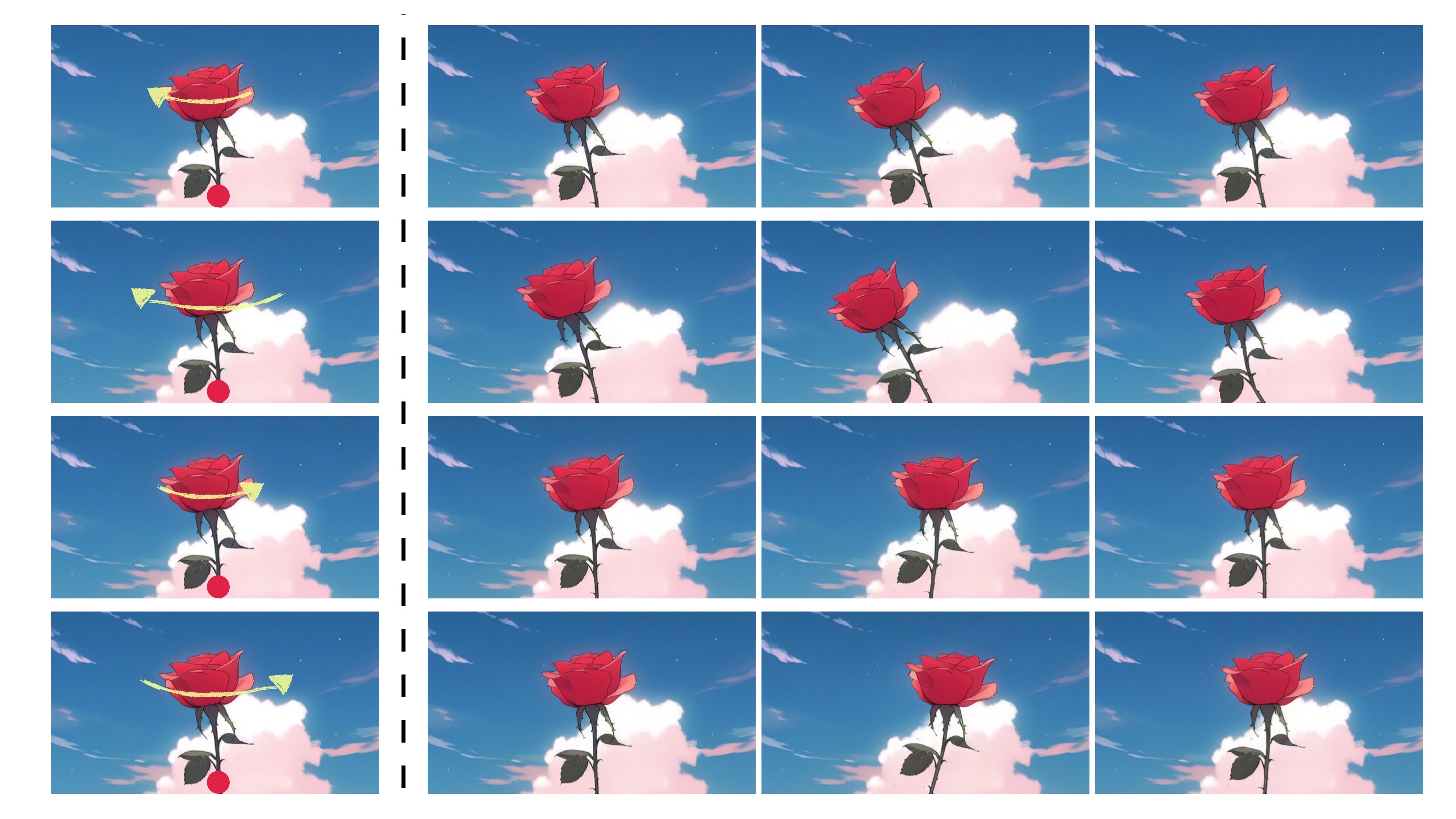}
    \caption{\textbf{Controllable Generation with Energy Stroke.} By applying energy strokes with varying directions and strengths, our method enables the convenient generation of diverse dynamic effects.}
    \label{fig:controllabel_generation2}
    \vspace{-10px}
\end{figure}

\begin{figure}
    \centering
    \includegraphics[width=1.0\linewidth]{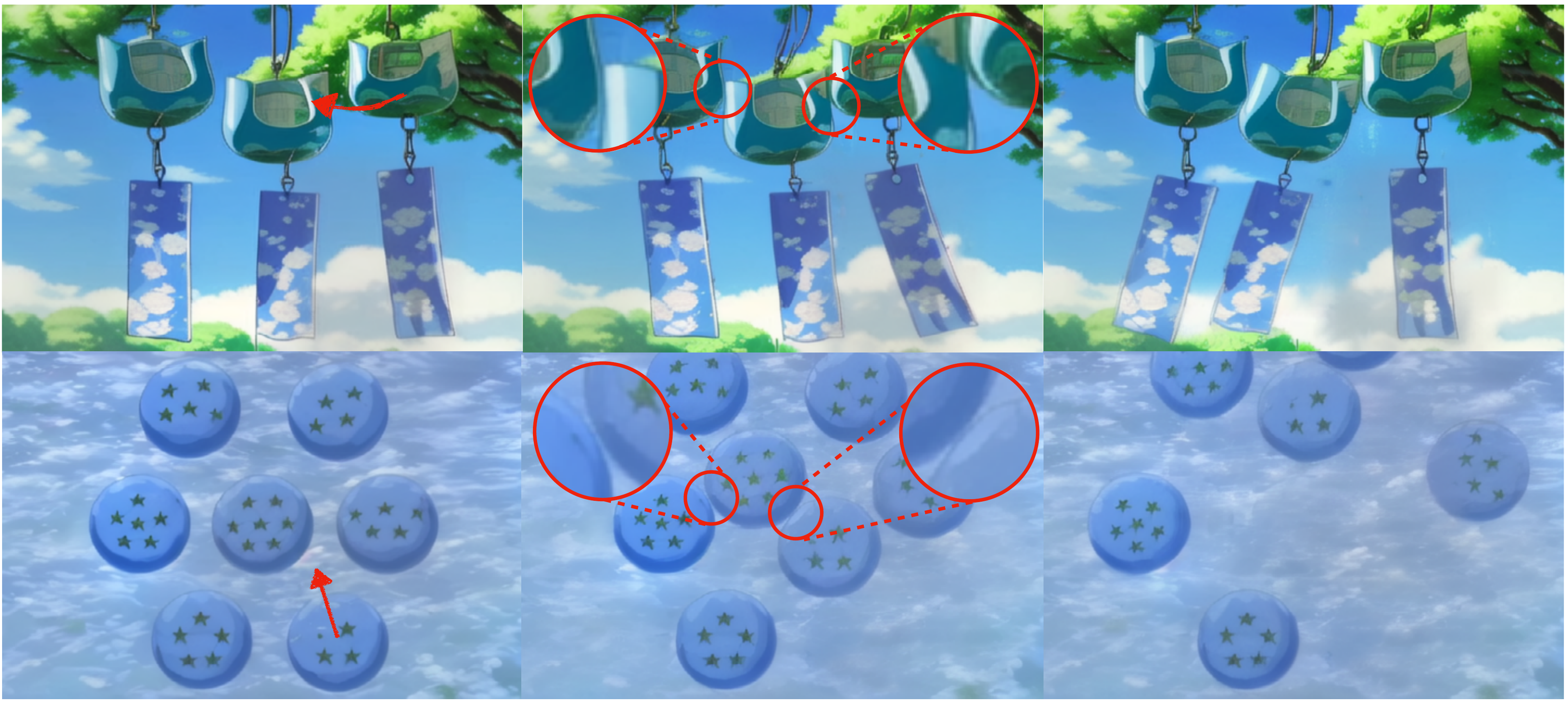}
    \caption{\textbf{Multi-Object Interaction.} Collisions are highlighted with red circles.}
    \label{fig:multiobject_interaction}
    \vspace{-10px}
\end{figure}
Our physics-based modeling approach offers flexibility in adjusting the properties of animated objects, such as stiffness, as shown in Fig.~\ref{fig:controllabel_generation}. Additionally, Fig.\ref{fig:controllabel_generation2} demonstrates how our method enables the convenient creation of diverse dynamic effects by applying energy brushes with varying settings.

\subsection{Multi-object Interaction}
Our approach inherently supports multi-object interactions, provided with simulated inter-object interactions, such as collision handling, shown in Figure~\ref{fig:multiobject_interaction}.

\section{Ablation Study}
We perform an ablation study to evaluate the effectiveness of our complementary dynamics enhancement module. As part of this evaluation, we conducted a user study where participants were asked, "Which video resembles real anime more?" We present the user preferences alongside the VideoScore \cite{he2024videoscore} metrics in Table~\ref{tab:ablation_study}. Although applying the dynamics enhancement results in a slight decrease in VideoScore metrics, the user study shows a clear preference (70.4\%) for the videos with dynamics enhancement, demonstrating this module’s contribution to improved anime stylization.
% As shown in Tab.~\ref{tab:ablation_study}, applying 
% To enable interactive animation, we introduce energy strokes that carry with flow particles exerting forces to the nearby vertices. The 
% \TODO{1. show why use Gaussian Blurred Sketch, 2. show the effectiveness of the interpolation module}

\section{More Comparison Results}
% \TODO{provide more qualitative comparison results}
In Fig.~\ref{fig:sup_qua1}, \ref{fig:sup_qua2} and ~\ref{fig:sup_qua3}, we provide additional qualitative comparison results with baseline methods, including Cinemo \cite{ma2024cinemo}, Drag Anything \cite{wu2025draganything}, DynamiCrafter \cite{xing2025dynamicrafter} and Motion-I2V \cite{shi2024motion}.

% \section{More Qualitative Results}
% We provide additional qualitative results in Fig.\TODO{}
\end{document}